\documentclass[twoside,twocolumn,9pt]{article}
\usepackage{extsizes}
\usepackage[super,sort&compress,comma]{natbib}
\usepackage[version=3]{mhchem}
\usepackage[left=1.5cm, right=1.5cm, top=1.785cm, bottom=2.0cm]{geometry}
\usepackage{balance}
\usepackage{mathptmx}
\usepackage{sectsty}
\usepackage{graphicx}
\usepackage{lastpage}
\usepackage[format=plain,justification=justified,singlelinecheck=false,font={stretch=1.125,small,sf},labelfont=bf,labelsep=space]{caption}
\usepackage{float}
\usepackage{fancyhdr}
\usepackage{fnpos}
\usepackage[english]{babel}
\addto{\captionsenglish}{%
  \renewcommand{\refname}{Notes and references}
}
\usepackage{array}
\usepackage{droidsans}
\usepackage{charter}
\usepackage[T1]{fontenc}
\usepackage[usenames,dvipsnames]{xcolor}
\usepackage{setspace}
\usepackage[compact]{titlesec}
\usepackage[colorlinks=true, citecolor=blue, linkcolor=blue, breaklinks=true]{hyperref}

\usepackage{amsmath}

\definecolor{cream}{RGB}{222,217,201}

\begin{document}

\pagestyle{fancy}
\thispagestyle{plain}
\fancypagestyle{plain}{
\renewcommand{\headrulewidth}{0pt}
}

\makeFNbottom
\makeatletter
\renewcommand\LARGE{\@setfontsize\LARGE{15pt}{17}}
\renewcommand\Large{\@setfontsize\Large{12pt}{14}}
\renewcommand\large{\@setfontsize\large{10pt}{12}}
\renewcommand\footnotesize{\@setfontsize\footnotesize{7pt}{10}}
\makeatother

\renewcommand{\thefootnote}{\fnsymbol{footnote}}
\renewcommand\footnoterule{\vspace*{1pt}%
\color{cream}\hrule width 3.5in height 0.4pt \color{black}\vspace*{5pt}}
\setcounter{secnumdepth}{5}

\makeatletter
\renewcommand\@biblabel[1]{#1}
\renewcommand\@makefntext[1]%
{\noindent\makebox[0pt][r]{\@thefnmark\,}#1}
\makeatother 
\renewcommand{\figurename}{\small{Fig.}~}
\sectionfont{\sffamily\Large}
\subsectionfont{\normalsize}
\subsubsectionfont{\bf}
\setstretch{1.125} 
\setlength{\skip\footins}{0.8cm}
\setlength{\footnotesep}{0.25cm}
\setlength{\jot}{10pt}
\titlespacing*{\section}{0pt}{4pt}{4pt}
\titlespacing*{\subsection}{0pt}{15pt}{1pt}

\fancyfoot{}
\fancyfoot[R]{\footnotesize{\hspace{2pt}\thepage}}
\fancyhead{}
\renewcommand{\headrulewidth}{0pt}
\renewcommand{\footrulewidth}{0pt}
\setlength{\arrayrulewidth}{1pt}
\setlength{\columnsep}{6.5mm}
\setlength\bibsep{1pt}

\makeatletter
\newlength{\figrulesep}
\setlength{\figrulesep}{0.5\textfloatsep}

\newcommand{\topfigrule}{\vspace*{-1pt}%
\noindent{\color{cream}\rule[-\figrulesep]{\columnwidth}{1.5pt}} }

\newcommand{\botfigrule}{\vspace*{-2pt}%
\noindent{\color{cream}\rule[\figrulesep]{\columnwidth}{1.5pt}} }

\newcommand{\dblfigrule}{\vspace*{-1pt}%
\noindent{\color{cream}\rule[-\figrulesep]{\textwidth}{1.5pt}} }

\makeatother

\twocolumn[
\newgeometry{onecolumn, centering, margin=4cm}
\vspace{1em}
\sffamily

\noindent\LARGE{\textbf{Effect of coat-protein concentration on the
    self-assembly of bacteriophage MS2 capsids around RNA}}
\\

\noindent\large{LaNell A. Williams,\textit{$^{a}$}
  Andreas Neophytou,\textit{$^{b}$}
  Rees F. Garmann,\textit{$^{c,d,e}$}
  Dwaipayan Chakrabarti,\textit{$^{b}$}
  Vinothan N. Manoharan,$^{\ast}$\textit{$^{c,a}$}} \\

   Self-assembly is a vital part of the life cycle of certain
   icosahedral RNA viruses. Furthermore, the assembly process can be
   harnessed to make icosahedral virus-like particles (VLPs) from coat
   protein and RNA \textit{in vitro}. Although much previous work has
   explored the effects of RNA-protein interactions on the assembly
   products, relatively little research has explored the effects of
   coat-protein concentration. We mix coat protein and RNA from
   bacteriophage MS2, and we use a combination of gel electrophoresis,
   dynamic light scattering, and transmission electron microscopy to
   investigate the assembly products. We show that with increasing
   coat-protein concentration, the products transition from well-formed
   MS2 VLPs to ``monster'' particles consisting of multiple partial
   capsids to RNA-protein condensates consisting of large networks of
   RNA and partially assembled capsids.  We argue that the transition
   from well-formed to monster particles arises because the assembly
   follows a nucleation-and-growth pathway in which the nucleation rate
   depends sensitively on the coat-protein concentration, such that at
   high protein concentrations, multiple nuclei can form on each RNA
   strand. To understand the formation of the condensates, which occurs
   at even higher coat-protein concentrations, we use Monte Carlo
   simulations with coarse-grained models of capsomers and RNA. These
   simulations suggest that the the formation of condensates occurs by
   the adsorption of protein to the RNA followed by the assembly
   of capsids. Multiple RNA molecules can become trapped when a capsid
   grows from capsomers attached to two different RNA molecules or when
   excess protein bridges together growing capsids on different RNA
   molecules. Our results provide insight into an important biophysical
   process and could inform design rules for making VLPs for various
   applications.

      \vspace{0.6cm}
\restoregeometry
]

\renewcommand*\rmdefault{bch}\normalfont\upshape
\rmfamily
\section*{}
\vspace{-1cm}


\footnotetext{\textit{$^{a}$~Department of Physics, Harvard University, Cambridge, MA 02138, USA.}}
\footnotetext{\textit{$^{b}$~School of Chemistry, University of Birmingham, Edgbaston, Birmingham B15 2TT, UK.}}
\footnotetext{\textit{$^{c}$~Harvard John A. Paulson School of Engineering and Applied Sciences, Harvard University, Cambridge, MA 02138, USA.}}
\footnotetext{\textit{$^{d}$~Department of Chemistry and Biochemistry, San Diego State University, San Diego, CA 92182, USA}}
\footnotetext{\textit{$^{e}$~Viral Information Institute, San Diego State University, San Diego, CA 92182 USA}}

\footnotetext{$^\ast$~vnm@seas.harvard.edu}










\section{Introduction}

For positive-strand RNA viruses to replicate, coat proteins must
assemble around the viral RNA to form new virus
particles.~\cite{caspar1962physical} Certain features of this assembly
process can be replicated \textit{in vitro}, in the absence of host-cell
factors.~\cite{fraenkel-conrat_reconstitution_1955,
  bancroft_formation_1967, hiebert_assembly_1968} For example,
virus-like particles (VLPs) can be assembled from solutions of the coat
protein and RNA of bacteriophage MS2. Wild-type MS2 particles have an
icosahedral capsid (triangulation number $T=3$, diameter about 30~nm)
containing one maturation protein and 178 coat proteins surrounding an
RNA strand with approximately 3600 nucleotides. By contrast, MS2 VLPs
that assemble \textit{in vitro} lack the maturation protein required for
infectivity. Nonetheless, they can adopt the same structure and size as
wild-type MS2 virus particles.~\cite{sugiyama1967ribonucleoprotein} This
result supports the premise that RNA virus assembly is driven by
free-energy minimization.

\begin{figure*}[!htb]
  \centering
  \includegraphics{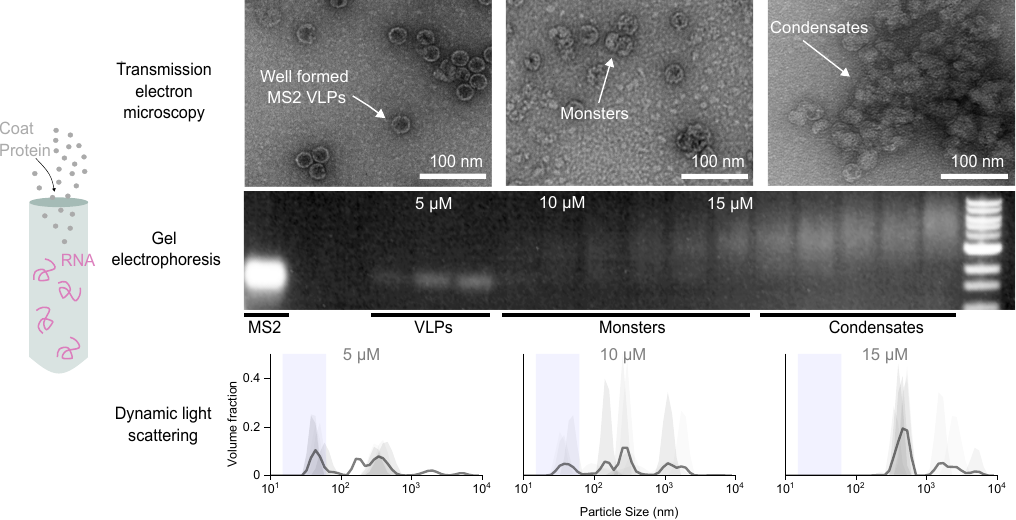}
  \caption{\label{fig:introfig_revised_300dpi} Overview of experiments
    and results. We mix MS2 coat protein with MS2 RNA to make a solution
    with 50~nM RNA concentration and varying coat-protein concentration.
    The transmission electron microscopy images, gel electrophoresis
    measurements (full image is shown in Fig.~\ref{fig:gel}), and
    dynamic light scattering results demonstrate a transition from
    well-formed MS2 VLPs to monster particles to RNA-protein condensates
    with increasing coat protein concentration. The main text and
    subsequent figures elaborate on all of these results.}
\end{figure*}

However, the assembly process itself and the conditions under which it
leads to well-formed structures are not yet well understood. In MS2,
most previous work on this question has focused on the role of specific
interactions between coat protein and the viral RNA.
Studies~\cite{beckett1988roles, peabody1993rna} on R17, a virus closely
related to MS2, have shown that the the overall yield of assembled VLPs
decreases if the RNA does not contain a sequence called the
translational operator that has a strong and specific affinity for coat
protein.~\cite{johansson1997rna} Nonetheless, assembly proceeds in the
absence of the operator, perhaps due to non-specific interactions
between the coat protein with the RNA~\cite{beckett1988roles}.
Therefore, specific RNA-protein interactions might affect the assembly
rate and yield but do not seem to be essential to the assembly process.

While these studies have established the relevance of RNA-protein
interactions to the assembly process, they did not directly reveal the
assembly pathway itself. More recent work involving interferometric
scattering microscopy, a technique that can image individual VLPs as
they form, shows that MS2 VLPs assemble by a nucleation-and-growth
pathway~\cite{garmann2019measurements} at near-neutral pH, salt
concentrations on the order of 100~mM, and micromolar coat-protein
concentrations. In this pathway, a critical nucleus of proteins must
form on the RNA before the capsid can grow to completion. The size of
the critical nucleus, estimated to be less than six coat-protein dimers,
is associated with a free-energy barrier. Taken together with the
previous experiments on the role of the RNA
sequence~\cite{beckett1988roles,peabody1993rna}, these results show that
MS2 assembly is a heterogeneous nucleation process, in which the
nucleation rate is likely controlled by two factors: RNA-protein
interactions and the coat-protein concentration.

The role of the protein concentration has been less well investigated
than the role of the RNA-protein interactions. The interferometric
scattering microscopy experiments~\cite{garmann2019measurements} showed
that very few VLPs are formed at low (1~$\mu$M) concentration of MS2
coat-protein dimers, while well-formed capsids form at higher
concentrations, and so-called ``monster'' particles, consisting of
multiple partially formed capsids on a single strand of RNA, form at an
even higher concentrations (several~$\mu$M). These results suggest that
the nucleation barrier, which controls the nucleation rate, depends
sensitively on the coat-protein concentration. At low concentration, the
nucleation rate is too small for capsids to form within the experimental
time frame; at high concentration, the nucleation rate is so high that
multiple nuclei can form on a single RNA strand, resulting in monster
particles. However, this study examined only a few protein concentrations,
and the experiments were performed at low RNA concentration relative to
protein.

Here we use bulk assembly experiments to determine the assembly products
of MS2 coat protein and MS2 RNA as a function of coat-protein
concentration. We characterize the assembly products using three
techniques: gel electrophoresis, dynamic light scattering (DLS), and
transmission electron microscopy (TEM). In comparison to the previous
study,\cite{garmann2019measurements} in which protein was in large
excess relative to RNA, our study examines a much wider range of
coat-protein concentrations, including ones near the stochiometric ratio
of coat protein to RNA. Furthermore, the three-pronged experimental
approach allows us to corroborate results and test hypotheses about how
the assembly products form. Gel electrophoresis and TEM provide
qualitative data that we use to determine the size and structure of the
assembly products, and DLS provides quantitative information about their
size distributions. With these methods, we show that as the coat-protein
concentration increases, the morphologies transition from well-formed
VLPs to monster particles to RNA-protein condensates consisting of large
networks of RNA and protein. These results are summarized in
Fig.~\ref{fig:introfig_revised_300dpi} and discussed in more detail in
Section~\ref{sec:results}. We explain these results with the aid of
simulations of coarse-grained models of capsomers and RNA.

\section{Results and Discussion}\label{sec:results}

\begin{figure*}[!htb]
  \centering
  \includegraphics{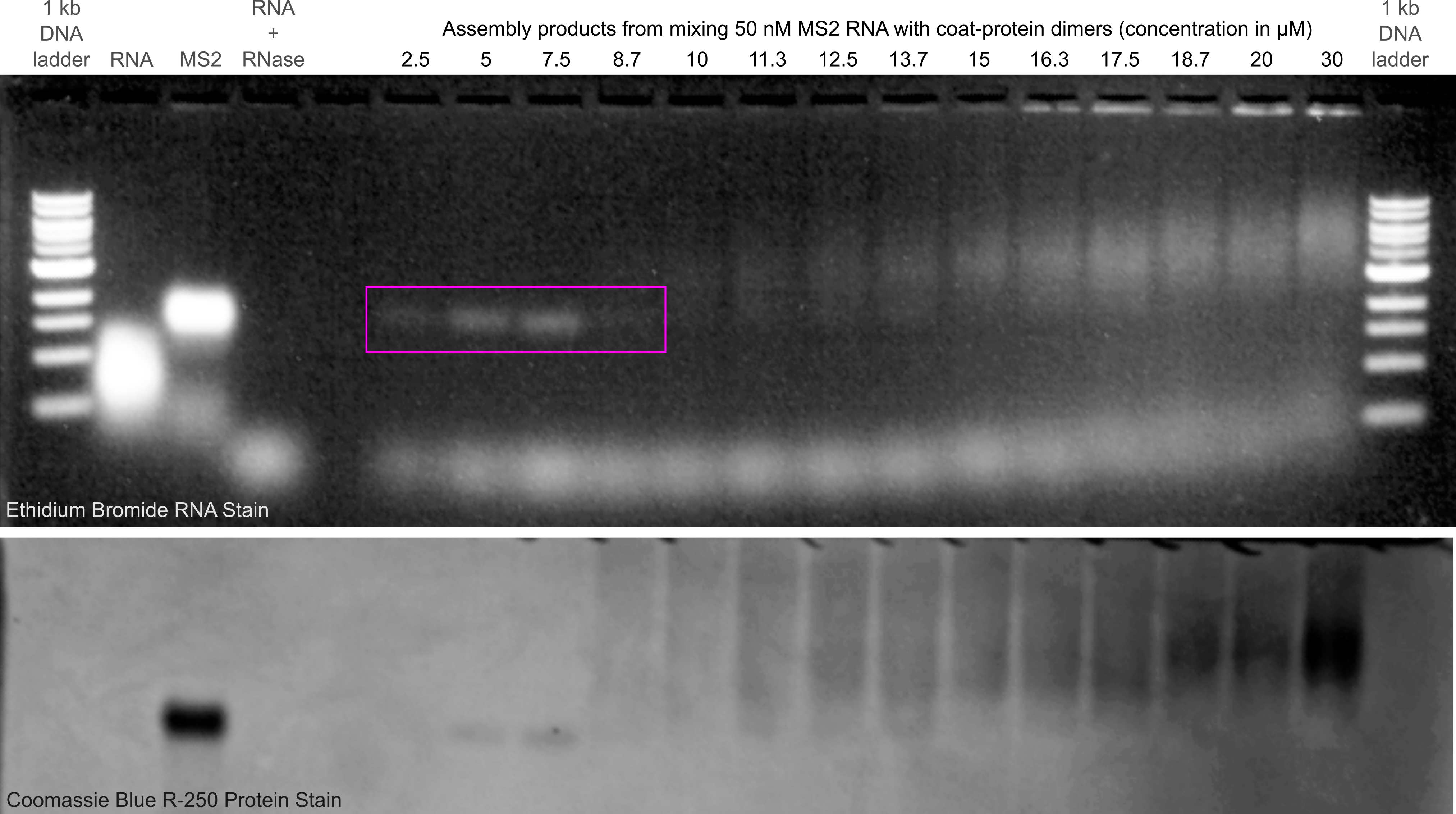}
  \caption{\label{fig:gel} Images of agarose gels used to characterize
    the assembly products. We first stain with ethidium bromide to
    detect the RNA (top image) and then stain with Coomassie Blue R-250
    to detect MS2 coat protein (bottom image). Lanes 1 (leftmost lane)
    and 20 (rightmost lane) show a DNA ladder. Lanes 2--4 show three
    controls: MS2 RNA (lane 2), wild-type MS2 capsids (lane 3), and MS2
    RNA treated with RNase (lane 4). We note that although usually MS2
    RNA runs at the same position as wild-type MS2, here we see that it
    runs farther, which may be because it has been exposed to
    contaminate RNAse from the neighboring lanes. The other lanes show
    the results of gel electrophoresis on samples prepared with
    coat-protein dimer concentrations ranging from 2.5~$\mu$M to
    30~$\mu$M. The region highlighted in purple shows that the amount of
    wild-type-sized products increases as dimer concentration increases
    from 2.5 to 7.5~$\mu$M and then decreases sharply at 8.7~$\mu$M.}
\end{figure*}

\subsection{Overview of experimental approach}
Briefly, our experimental procedure consists of combining 50~nM MS2 RNA
with purified MS2 coat-protein dimers at concentrations ranging from 2.5
to 30~$\mu$M (see Section~\ref{sec:methods} for full details). For
reference, a full VLP has an icosahedral capsid with a triangulation
number of 3 ($T=3$), corresponding to 180 coat proteins or 90
coat-protein dimers. At 50~nM RNA concentration, a coat-protein dimer
concentration of 5~$\mu$M therefore corresponds approximately to the
stoichiometric ratio of coat proteins to RNA in a full VLP. We work with
dimer concentrations instead of monomer concentrations because MS2 coat
proteins are thought to be dimerized in
solution.~\cite{lalwani2021understanding} After mixing the RNA and coat
protein, we then wait 10~min to allow assembly to occur. We chose this
time scale to be much larger than the assembly time observed in previous
experiments\cite{garmann2019measurements} at low protein concentration.
These experiments showed that at 2~$\mu$M protein dimer concentration
(lower than the lowest concentration in the current study, 2.5~$\mu$M),
capsids assembled in about 1--2~min. After 10~min, we add RNase to
digest any excess MS2 RNA that is not encapsidated. We then characterize
the resulting assembly products with gel electrophoresis, DLS, and TEM
(see Section~\ref{sec:methods}).

\subsection{Results from gel electrophoresis}
We first qualitatively characterize the size and composition of the
assembly products using agarose gel electrophoresis. We use both
ethidium stain to detect RNA and Coomassie stain to detect coat protein
in our samples. For comparison, we also characterize wild-type MS2, MS2
RNA, and digested MS2 RNA (see Section~\ref{sec:methods}).

The most striking feature of the gel is a band that runs at the same
position as wild-type MS2 but with a brightness that increases from 2.5
to 7.5~$\mu$M coat-protein dimers and then suddenly decreases at
8.7~$\mu$M (see highlighted region in Fig.~\ref{fig:gel}). We interpret
this increase and sudden decrease as follows. Near the stoichiometric
ratio (approximately 5~$\mu$M dimers to 50~nM RNA), well-formed VLPs
assemble, with more VLPs forming at higher protein concentration. Above
7.5~$\mu$M, the sharp decrease in brightness indicates that far fewer
well-formed MS2 VLPs assemble. Instead, as indicated by the spreading of
the band toward to the upper part of the gel, the assembly products at
dimer concentrations greater than 7.5~$\mu$M are larger than the
wild-type particles. These assembly products appear in both gels in
Fig.~\ref{fig:gel}, indicating that they contain both RNA and protein.

We also see that at coat-protein dimer concentrations higher than
7.5~$\mu$M, the intensity of the diffuse band increases with increasing
concentration (Fig.~\ref{fig:gel}). The increase in brightness and
change in the center position of this band suggest that the amount of
large assembly products increases at the expense of the wild-type-sized
products. At 15~$\mu$M, the diffuse band no longer overlaps with the
band corresponding to wild-type-size VLPs. For dimer concentrations
beyond 15~$\mu$M, some of the assembly products are so large that they
are trapped near the top of the agarose gel.

The transition from a bright to a diffuse band might represent a
transition from well-formed VLPs to either malformed structures or
aggregates of capsids. The gels by themselves cannot confirm
either hypothesis, since they reveal only that the assembly products all
contain RNA and that they increase in size with increasing coat-protein
concentration. We therefore turn to dynamic light scattering and
transmission electron microscopy experiments, as described below.

\subsection{Results from dynamic light scattering}

To quantify the sizes of the assembly products, we use DLS with
numerical inversion methods. These methods yield the size distributions
of assembly products in both number and volume bases (see
Section~\ref{sec:methods}).

\begin{figure}[!hb]
  \centering
  \includegraphics{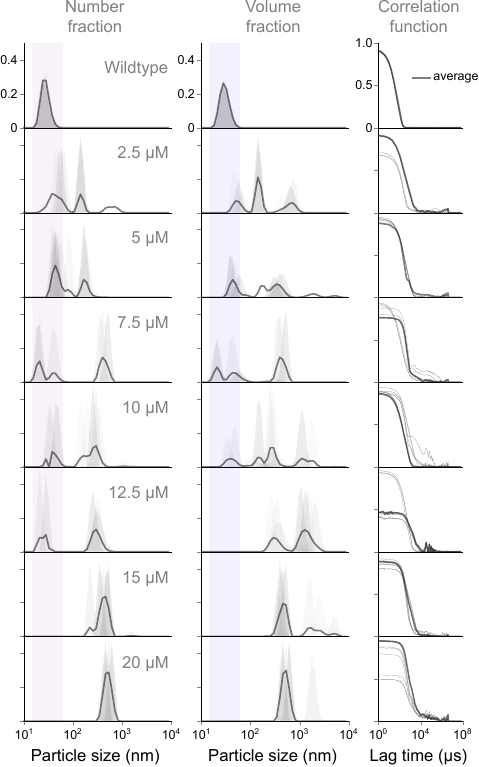}
  \caption{\label{fig:dls} Plots of size distributions of wild-type MS2
    virus particles and VLPs assembled \textit{in vitro} at 50~nM
    concentration of free RNA and varying coat-protein dimer
    concentrations. The distributions are inferred from dynamic light
    scattering measurements. The first column shows the size
    distribution on a number basis, the second column shows the size
    distribution on a volume basis, and the third column shows the
    measured autocorrelation functions. Light gray peaks in the
    distributions show the results from eight individual experiments.
    Dark gray peaks show the results inferred from the average
    autocorrelation function. The purple and blue shaded regions show
    the range of particle sizes consistent with the wild-type size
    distribution. The autocorrelation functions for each individual
    measurement are shown in light grey in the plots at right, and the
    average is shown in dark gray.}
\end{figure}

At coat-protein dimer concentrations 7.5~$\mu$M and below, we observe in
both the number and volume distribution a peak at or near the size of
wild-type MS2 particles (see shaded bands in Fig.~\ref{fig:dls}; we
expect some variation in the location of this peak because the inversion
of the autocorrelation function is sensitive to noise). This peak is
accompanied by peaks at larger sizes, unlike the size distribution for
wild-type MS2, which consists of only one peak. At coat-protein dimer
concentrations above 7.5~$\mu$M, the peak corresponding to the size of
wild-type MS2 particles decreases until it disappears (in the
volume-basis distributions) at 12.5~$\mu$M. At concentrations of 15 and
20~$\mu$M, we observe a single peak corresponding to much larger
assembly products. Overall, we observe that the average size of the
assembly products increases with increasing protein concentration
(Fig.~\ref{fig:dls}).

The DLS data support our interpretation of the gel-electrophoresis data.
Specifically, both the DLS and gel data show that the proportion of VLPs
with sizes corresponding to the wild-type size decreases with
concentration above 7.5~$\mu$M, whereas only larger products form at
high concentration. The DLS data additionally show that the size of
these larger products is on the order of several hundred nanometers.

However, the DLS data also show peaks corresponding to particles larger
than wild-type at concentrations less than 10~$\mu$M. We do not see
evidence of such particles in the gel data. These peaks may correspond
to weakly-bound clusters of well-formed MS2 VLPs that are observable in
the DLS experiments but fall apart during gel electrophoresis (see
Fig.~\ref{fig:gel}). Because DLS does not provide any structural
information, we turn to TEM to characterize the structures of the
assembly products.

\subsection{Transmission electron microscopy (TEM) experiments}

\begin{figure*}[!htb] 
  \centering
\includegraphics[width=\textwidth]{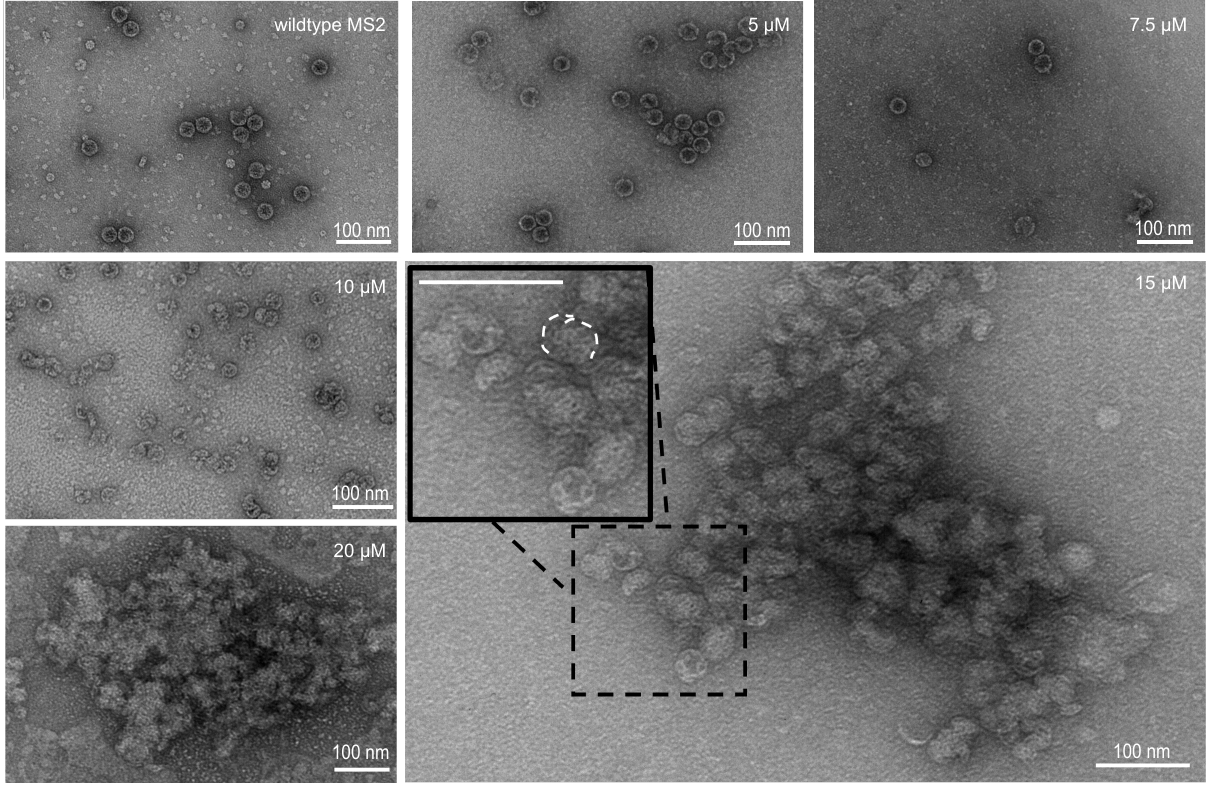}
\caption{\label{fig:temfig_editable_revised_300dpi} TEM images from
  negatively stained samples of wild-type MS2 particles and products of
  assembly at varying coat-protein dimer concentrations. At
  concentrations less than 10~$\mu$M, most particles have the shape and
  size of wild-type capsids. At higher concentrations, we observe
  clusters of partially formed capsids that increase in size with
  concentration. The dotted line in the inset of the 15~$\mu$M image
  shows the outline of one such partial capsid. All scale bars are
  100~nm.}
\end{figure*}

TEM images of negatively stained samples show that most of the assembly
products at dimer concentrations 7.5~$\mu$M and below are well-formed
MS2 VLPs (Figs.~\ref{fig:temfig_editable_revised_300dpi}
and~\ref{fig:farviewcondensates_revised_300dpi}), with some malformed
VLPs and clusters of MS2 VLPs, consistent with the larger sizes present
in the DLS-derived size distributions. At a concentration of 10~$\mu$M,
we observe malformed particles that consist of partially formed capsids.
These structures are similar to the so-called ``monster'' particles
observed in turnip-crinkle-virus assemblies~\cite{sorger1986structure}
and, more recently, in MS2 assembly
experiments.~\cite{garmann2019measurements} At concentrations above
15~$\mu$M we observe what appear to be large aggregates of partially
formed capsids (Figs.~\ref{fig:temfig_editable_revised_300dpi}
and~\ref{fig:farviewcondensates_revised_300dpi}). These structures are
micrometer-sized, comparable to the sizes seen in the DLS distributions
(Fig.~\ref{fig:dls}).

\subsection{Discussion of experimental results}

Our measurements show that coat-protein concentration plays an important
role in the morphology of the assembly products of MS2 RNA and coat
protein. At low coat-protein dimer concentrations (less than
7.5~$\mu$M), gel electrophoresis, DLS, and TEM all point to the
formation of MS2 VLPs that are of the same size as wild-type MS2. These
structures appear to be well-formed, consistent with previous
studies.~\cite{sugiyama1967ribonucleoprotein} At higher concentrations
(between 7.5 and 10~$\mu$M), we observe monster particles consisting of
a few partial capsids and RNA. While we cannot determine from the data
whether the monster particles form around a single or multiple strand of
RNA, the monster particles have been observed in previous experiments on
MS2 assembly,\cite{garmann2019measurements} and interferometric
scattering measurements indirectly show that they can grow around a
single RNA strand. At an even higher concentration (12.5~$\mu$M),
results from gel electrophoresis, DLS, and TEM point to the formation of
large structures several hundred nanometers in size and containing many
partial capsids and RNA.

Whereas the observation of well-formed VLPs and even monsters is
consistent with previous studies on MS2, the observation of large
structures at high protein concentrations has not, to our knowledge,
been studied in detail. Large structures have been observed in the
assembly of viral coat proteins around functionalized gold
nanoparticles, but these structures are found at low protein
concentrations.~\cite{malyutin_budding_2013} In other viruses, large
aggregates have been observed under conditions of strong
interactions.~\cite{garmann2014assembly, garmann2016physical} Here,
however, the formation of the large structures occurs at the same buffer
conditions (apart from coat-protein concentration) as those used to
assemble well-formed VLPs.

The large structures are interesting not only because they contain many
partially formed capsids, but also because they contain RNA, as shown by
our gel electrophoresis measurements. We term these structures
``condensates'' because, like other biological structures that bear this
name,\cite{alshareedah_interplay_2019, guillen-boixet_rna-induced_2020}
they are self-organized and contain both RNA and protein.

The formation of the condensates points to a more complex pathway than
the one that appears to be operative at lower protein concentrations.
The well-formed VLPs at low concentrations and monster particles at
intermediate concentrations can be explained in terms of a nucleated
pathway~\cite{garmann2019measurements} in which the nucleation rate
increases with protein concentration. The monster particles, which
consist of multiple partial capsids, can form when more than one
nucleation event happens on a single RNA strand; indeed, we expect that
the probability of multiple nucleation events should increase with the
protein concentration. However, the size of the condensates (and their
fluorescence in the gel assays under RNA staining) suggests that
they contain multiple RNA molecules.  It is not immediately obvious how
a high nucleation rate could lead to multiple RNA molecules becoming
trapped between partial capsids.

One hypothesis is that condensate formation is driven primarily by
aggregation of coat proteins. If the aggregation of the coat proteins
were rapid, it is possible that the RNA molecules could be trapped
inside the aggregate. However, gel electrophoresis, DLS, and TEM
experiments show no evidence of coat-protein aggregation in the absence
of RNA, even at 15~$\mu$M dimer concentration.

\begin{figure}[htb]
\includegraphics[width=\columnwidth]{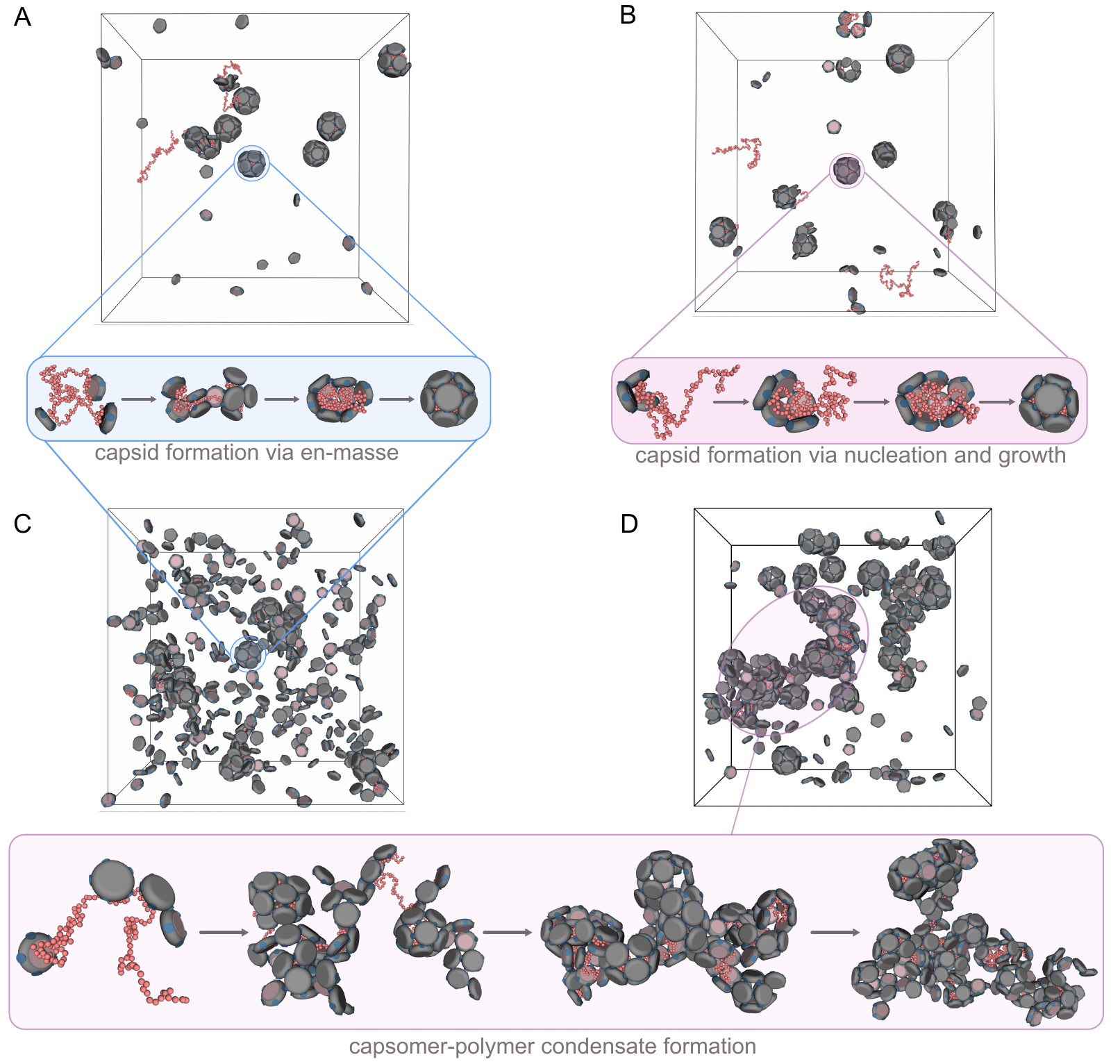}
\caption{\label{fig:simulation} Representative snapshots from Monte
  Carlo simulations of capsomer-and-polymer systems with (A,B) a 12:1
  ratio of capsomer to polymer and (C,D) a 50:1 ratio. The volumes of
  the simulated systems are all identical (see
  Section~\ref{sec:methods}). Each capsomer is modeled as a hard disk
  with five sticky patches on its rim that mediate capsomer-capsomer
  interactions, and a large sticky patch on its face that mediates
  capsomer-polymer interactions. The reduced temperatures and relative
  energies of capsomer-capsomer and capsomer-polymer interactions are
  chosen such that capsid formation in the low capsomer:polymer ratio
  systems proceeds either by an (A) en-masse pathway or (B) a
  nucleation-and-growth pathway, as highlighted by the representative
  configurations along capsid-forming trajectories for each simulation
  (see shaded subpanels). When the interactions favor en-masse assembly
  at a low capsomer:polymer ratio, increasing the capsomer:polymer ratio
  leads to the formation of discrete capsids (C). But when the
  interactions favor nucleation and growth at a low capsomer:polymer
  ratio, increasing the capsomer:polymer ratio leads to the formation of
  extended structures mediated by capsomer-capsomer interactions (D).
  These structures contain multiple polymer chains and resemble the
  RNA-protein condensates found in the experiments. The subpanel below
  panel D shows snapshots of the condensate as it assembles. The number
  of chains in the condensate increases from 1 to 7 in this trajectory.}
\end{figure}

\subsection{Coarse-grained modeling}

We turn to coarse-grained modeling to gain insight into the potential
pathways. In the simulations, we model the capsomers as patchy hard
disks and the RNA as a polymer with a length of approximately 14 times
the diameter of a fully formed capsid (see Fig.~\ref{fig:simulation} and
Section~\ref{sec:methods}), such that each polymer can be encapsidated
by 12 capsomers. In contrast to previous coarse-grained simulations that
focused on capsomer assembly around a single
polymer\cite{elrad2010encapsulation, perlmutter2013viral,
  perlmutter2014pathways} -- or, at most, a few
polymers\cite{zhang2014topological} -- our simulations include larger
numbers of polymers (10 to 30 within the volume of the simulated system)
and therefore allow for the possibility of condensate formation.
Nonetheless, the simulated system is simplified from the experimental
system in several ways: whereas an MS2 VLP consists of 90 coat-protein
dimers, yielding a $T=3$ structure, a complete capsid in the simulation
consists of 12 pentamers; whereas MS2 can adopt intricate secondary and
tertiary structures, the simulated polymer does not; whereas the MS2
protein subunits are not rigid bodies but instead can stretch, which may
help promote assembly,\cite{panahandeh_equilibrium_2018,
  panahandeh_how_2020} we ignore this elastic-energy contribution in
our simulation. We choose to simplify these features so that the
simulation can give insight into the simplest potential pathways leading
to condensate-like structures.

We perform two sets of initial simulations in which the strength of
capsomer-polymer interactions is tuned so that, in one set, capsids
nucleate and grow at low capsomer concentrations, and, in the other set,
capsids assemble ``en masse.''\cite{elrad2010encapsulation} In the
en-masse pathway, many capsomers first bind to the polymer in a
disordered arrangement and then form a capsid. In both cases, the
simulated system contains 10 polymers. At low capsomer concentrations,
both sets of simulations show the assembly of well-formed capsids
containing the polymer, as expected (Fig.~\ref{fig:simulation}A and~B).
But at high capsomer concentrations, we observe differences. Whereas the
system that follows an en-masse pathway at low concentrations still
forms capsids at higher concentrations (Fig.~\ref{fig:simulation}C), the
system that follows nucleation and growth at low concentrations
assembles into large networks of polymers and partial capsids
(Fig.~\ref{fig:simulation}D), resembling the condensates seen in the
experiments. To ensure that the formation of the condensates is not an
artifact of the finite size of the system, we perform additional
simulations in the nucleated regime with 30 polymers and find the same
result. These simulations show that the formation of condensate-like
structures does not require strong protein-RNA interactions; that is,
condensate-like structures can occur under interactions that, at lower
protein concentrations, lead to the nucleation and growth of full
capsids.

To support these visual observations, we calculate the distribution of
the average size of the clusters containing both polymers and capsomers
over a fixed number of configurations in each of the simulations
(Fig.~\ref{fig:Simulation_DLS}). These distributions qualitatively
resemble those obtained by DLS (Fig.~\ref{fig:dls}). Specifically, we
find that both sets of simulations have a single peak centered at a
cluster size of 12 for low capsomer:polymer ratios, corresponding to the
formation of complete and dispersed capsids. Both sets of simulations
also show a peak centered at a cluster size of 20--30, corresponding to
the formation of monster particles. However, at high
capsomer:polymer ratios we see that the average cluster size is
approximately 20 for the system that followed an en-masse pathway at
lower concentrations, reflecting the formation of only dispersed capsids
and monsters, with no condensate-like structures. By contrast, we
observe a broad peak representing cluster sizes of hundreds for the
system that followed nucleation-and-growth at lower concentrations,
reflecting the formation of polymer-capsomer condensates.

To obtain insights into the formation of condensates, we perform
additional simulations with polymer chains that are 33\% shorter. We
work with the same system that shows nucleation and growth at low
concentrations and condensate formation at higher concentrations with
the standard-length polymer. With the shorter polymer, we find that the
condensates no longer form. Instead, discrete capsids and monsters now
assemble instead (Fig.~\ref{fig:Simulation_short_poly}). However, for
both polymer lengths, we observe that the capsomers first bind to the
RNA, and then form partial or full capsids (see shaded subpanels in
Figs.~\ref{fig:simulation} and~\ref{fig:Simulation_short_poly}). The
observed assembly trajectories appear to follow an en-masse pathway --
and indeed, with short polymers, we directly see an en-masse-type
assembly (blue subpanel in Fig.~\ref{fig:Simulation_short_poly}). In
contrast to an en-masse pathway that occurs at low capsomer
concentrations, here the adsorption of the capsomers to the polymer is
not driven by strong capsomer-polymer interactions but instead by the
abundance of capsomers. For the standard-length polymers, however, this
assembly pathway leads to condensate-like structures, whereas for
shorter polymers it leads to dispersed capsids or monsters.

The simulations show that the large networks observed for the
standard-length polymers consist of multiple polymer strands (see bottom
shaded subpanel in Fig.~\ref{fig:simulation}). There are several ways in
which multiple polymers can become trapped in the same large structure:
a capsid might assemble around two polymer strands, or capsomers might
connect together or bridge partial capsids on two separate polymers, for
example. We expect the chances of both of these events occuring to
increase with the size of the polymer, since the distance between
segments on two different polymer chains should decrease with the
polymer size. These and similar mechanisms may explain why we observe
the condensate-like structures form for the standard-length polymers but
not the shortened ones.

We note also that although we expect the nucleation barrier to become
smaller as the capsomer concentration increases, the formation of
capsids still occurs heterogeneously; that is, we do not observe the
formation of empty capsids. In a typical phase separation, the
disappearance of the nucleation barrier is associated with spinodal
decomposition, which can also lead to the formation of extended
structures\cite{lu_gelation_2008}. Here, however, the heterogeneous
nature of condensate formation suggests a pathway different from
spinodal decomposition, since capsomers must first adsorb to the polymer
before capsids can form.  The absence of condensates from simulations for
shorter polymers also points to the importance of capsomer-polymer
associations in the pathway.

\section{Conclusions}

Our experiments and simulations show that when a nucleation-and-growth
pathway for capsid assembly is operative at low protein concentrations,
monster particles and RNA-protein condensates can form at higher
concentrations. The formation of the monster particles can be explained by
the increase in nucleation rate with increasing protein concentration.
When the timescale of nucleation is short compared to the time for a
nucleus to grow into a full capsid, multiple nuclei can form on the same
RNA strand. When these nuclei grow, they tend to form partial capsids
because other partial capsids on the same RNA can block their growth.

The condensates, however, appear to arise from a more complex pathway.
The pathway suggested by simulations involves proteins first attaching
to the RNA, then starting to assemble. At the high protein
concentrations that lead to condensate formation, proteins can bridge
together capsids that are growing on different RNA strands; also,
capsids may assemble around portions of two different RNA molecules.
These and related mechanisms would explain how, as seen in our
experimental results, the condensates grow so large and why they contain
multiple RNA strands and many partial capsids. Other hypotheses, such as
coat-protein aggregation or spinodal decomposition, do not account for
all of our results.

There remain a few questions to be resolved in future studies. One
question is how the RNA is spatially distributed in the condensates, and
whether the mechanisms by which multiple RNA strands become trapped in
the condensate, as observed in the simulations, are operative in the
experiments. Another question is what happens at concentrations between
those at which well-formed capsids form and monster particles form. At
these concentrations, DLS measurements show evidence for some structures
that are larger than single capsids. Because TEM data show that
most structures at these concentrations are not malformed, one
possibility is that the DLS measurements are detecting small clusters of
well-formed capsids. The driving force for the formation of these
clusters is not clear, but they might arise when a single RNA molecule
spawns multiple nuclei that each form a full (or nearly full) capsid. In
this situation, the RNA would connect the capsids into a ``multiplet''
structure.~\cite{garmann2014assembly} It is still not clear why the gel
measurements do not show evidence for such structures, however.
Fluorescent microscopy experiments could help resolve this question and
the aforementioned ones as well.

Our work might also inform models of the assembly pathway, particularly
those based on the law of mass action,~\cite{zlotnick94, zandi06:cnt,
  morozov09:assem, zandi96van, vanderholst18} in which the
concentration of coat proteins plays a critical role. Further
experiments that quantify how the nucleation rate depends on the
coat-protein concentration would help connect these models to the
morphological observations we present here. From a more practical
perspective, our work helps establish constraints on concentration for
the production of MS2 VLPs. Such VLPs are used to encapsulate materials
for drug delivery~\cite{kovacs07, galaway13, hartman2018quantitative}
and to display epitopes for vaccines.~\cite{peabody2021rna,
  peabody2008immunogenic}

\section{Methods and Materials}\label{sec:methods}

All materials were used as received.  Buffers were prepared as follows:
\begin{itemize}
\item Assembly buffer: 42 mM Tris, pH 7.5; 84 mM NaCl; 3 mM acetic acid, 1 mM EDTA
\item TNE buffer: 50 mM Tris, pH 7.5; 100 mM NaCl, 1 mM EDTA
\item TE buffer: 10 mM Tris, pH 7.5; 1 mM EDTA
\item TAE buffer: 40 mM Tris-acetic acid, pH 8.3; 1 mM EDTA
\end{itemize}

\subsection{Virus growth, cultivation, and storage}\label{sec:methodsgrowth}
We purify wild-type bacteriophage MS2 as described by Strauss and
Sinsheimer.~\cite{strauss1963purification} In brief, we grow MS2 virus
particles by infecting \textit{E.\@ coli} strain C3000 in minimal LB
Buffer, and we remove \textit{E.\@ coli} cell debris by centrifugation
at 16700$g$ for 30~min. We then use chloroform (warning: hazardous; use
in fume hood) extraction to purify the solute containing the virus. We
extract the purified virus particles by density gradient centrifugation
in a cesium chloride gradient. We store the purified virus at
4~\textdegree{}C at a concentration of 10$^{11}$ plaque-forming units
(pfu) in Tris-NaCL-EDTA or TNE buffer (50 mM Tris, 100 mM NaCL, 5 mM
EDTA) at pH 7.5. We determine the concentration of virus by
UV-spectrophotometry (NanoDrop 1000, Thermo Scientific) using an
extinction coefficient of 8.03~mL/mg at 260~nm.

\subsection{Coat-protein purification and storage}\label{sec:methodscp}

We purify MS2 coat-protein dimers following the method of Sugiyama,
Herbert, and Hartmant.~\cite{sugiyama1967ribonucleoprotein} Wild-type
bacteriophage MS2 is suspended in glacial acetic acid (warning:
hazardous; use in fume hood with appropriate personal protective
equipment) for 30 min to denature the capsid, separate it into protein
dimers, and precipitate the RNA. We then centrifuge the sample at
10000$g$ and collect the supernatant, which contains coat-protein
dimers. We filter out the glacial acetic acid with 20~mM acetic acid
buffer through 3-kDa-MWCO sterile centrifugal filters (Millipore Sigma,
UFC500324) five times. This process removes the glacial acetic acid to
prevent further denaturing of the coat-protein dimers. We then determine
the concentration of our coat-protein dimers by measuring the absorbance
with the Nanodrop Spectrophotometer (Thermo Fisher) at 280~nm. We store
the MS2 coat protein at 4~\textdegree{}C in a 20~mM acetic acid buffer.
We measure the absorbance at 260~nm to detect residual RNA. In our
experiments, we use only purified protein with an absorbance ratio
(protein:RNA) above 1.5 to avoid RNA contamination.

\subsection{RNA purification and storage}\label{sec:methodsrna}

We purify wild-type MS2 RNA using a protocol involving a Qiagen RNeasy
Purification Kit Mini (Qiagen, 7400450). We take 100~$\mu$L of MS2
stored in TNE buffer and mix with 350~$\mu$L of buffer RLT (a lysis
buffer) to remove the coat-protein shell. We add 250~$\mu$L of ethanol
to our sample and mix to precipitate the RNA. We then transfer our
sample to a 2~mL RNeasy Mini spin column (provided by the Qiagen
Purification Kit) that is placed in a collection tube. We then
centrifuge at 10000$g$ for 15~s and discard the flow-through. We add
500~$\mu$L of buffer RPE (to remove traces of salts) to the spin column
and centrifuge for 15~s at 10000$g$. We discard the flow-through. We
then add 500~$\mu$L of buffer RPE once more to the spin column and
centrifuge for 2~min at 10000$g$. We place the spin column upside down
into in a fresh 1.5~mL collection tube (provided in the purification
kit) to collect the RNA trapped in the spin column. We add 50~$\mu$L of
TE buffer to the spin column and centrifuge at 10000$g$ for 1~min to
collect the RNA. We measure the RNA concentration using a Nanodrop
spectrophotometer by measuring the absorbance at 260~nm and using an
extinction coefficient of 25.1~mL/mg. We store the purified MS2 RNA at
-80\textdegree{}C in Tris-EDTA (TE) buffer at neutral pH (7.5).

\subsection{RNA and coat-protein bulk assembly experiments}\label{sec:methodsbulk}

For assembly experiments, we mix wild-type MS2 RNA genome at a
concentration of 50~nM with varying concentrations of MS2 coat-protein
dimers ranging from 2.5~$\mu$M to 30~$\mu$M. We leave the mixtures at
room temperature (21~\textdegree{}C) for 10~min. Afterward, we add 10~ng
of RNase A to the sample and wait 30 min. We then characterize the
assembled virus-like particles using gel electrophoresis, dynamic light
scattering (DLS), and transmission electron microscopy (TEM).

\subsection{Gel electrophoresis and analysis}\label{sec:methodsgel}

For gel electrophoresis experiments, we mix 15~$\mu$L of sample with
4~$\mu$L of glycerol and load into a 1\% agarose gel in assembly buffer
consisting of 5 parts Tris-NaCL-EDTA (TNE) buffer (50~mM Tris, 100~mM
NaCl, 10~mM EDTA, pH 7.5) to 1 part 20~mM acetic acid buffer. We use
Ethidium Bromide (EtBr; warning: hazardous; use in fume hood with
appropriate personal protective equipment) to stain the RNA and to
detect the presence of MS2 RNA. We use Coomassie Blue R-250 to detect
the presence of MS2 coat protein. The combination of these staining
methods allow us to confirm the presence of both MS2 RNA and MS2 coat
protein within the resulting assemblies. We place three control samples
in lanes 2 through 4 that include MS2 RNA at 50~nM concentration (lane
2), wild-type MS2 at 50~nM concentration (lane 3), and 50~nM
concentration of digested MS2 RNA genome (lane 4) resulting from the
addition of RNase A. These controls allow us to compare the sizes of our
assembly products to systems of known sizes. We can also determine
whether the samples consist of MS2 VLPs formed during assembly or excess
strands of MS2 RNA. We place our assembly products in lanes 6 through
19. These samples are loaded and run at 21~\textdegree{}C at 100~V for
40~min and visualized using a Biosystems UV Imager (Azure, AZ1280).

\subsection{Dynamic light scattering (DLS) and analysis}\label{sec:methodsdls}
We use dynamic light scattering (Malvern ZetaSizer Nano ZS by Malvern
Panalytical) to determine the size distribution of particles that
assemble at 50~nM MS2 RNA concentration and coat-protein dimer
concentrations of 2.5, 5, 7.5, 10, 12.5, 15, and 20~$\mu$M. In each case
the samples are treated with RNase as described previously. We also
characterize the wild-type virus for comparison. We determine the size
distributions using the regularization inversion method provided by the
instrument software.~\cite{zeta07}

\subsection{Transmission electron microscopy (TEM) and analysis}\label{sec:methodstem}

For transmission electron microscopy, we negatively stain samples that
have been assembled in bulk at coat-protein dimer concentrations of 5,
7.5, 10, 15, and 20 $\mu$M and treated with RNase A. We stain with 2\%
aqueous uranyl acetate (warning: hazardous; use with appropriate
personal protective equipment) on 200 mesh carbon-coated copper TEM
grids (Polyscience, TEM-FCF200CU), then image with a Hitachi 7800 TEM
located at the Center for Nanoscale Systems at the Science and
Engineering Complex (CNS-SEC) at Harvard University. Images are taken at
20, 50, and 100~kV.

As a control, we mix 15~$\mu$M MS2 coat-protein dimers in assembly
buffer. This control is done to ensure that capsid-like or VLP-like
structures do not form in the absence of MS2 RNA.

\subsection{Coarse-grained model for capsid assembly}

We developed a patchy particle model for the capsomers interacting with
a polymer chain, which was used to model the RNA, to investigate their
assembly. A capsid is constructed from 12 subunits, each having $C_{5v}$
symmetry, where the center of each subunit sits on the vertex of an
icosahedron.\cite{wales2005energy, fejer2009energy,
  johnston2010modelling, perlmutter2014pathways}

\subsubsection{Capsomer-Capsomer Interactions.}
We coarse-grain the capsomeric building blocks as oblate hard
spherocylinders (OHSCs) decorated with five identical circular patches
conforming to $C_{5\text{v}}$ symmetry. See Fig.~\ref{fig:patchymodel}
for a schematic illustration of the model capsomer. For hard oblate
spherocylinders, which were previously used as a model system to
investigate the phase behavior of discotic liquid
crystals,~\cite{cuetos2008columnar} the surface is defined by the points
at a distance $L/2$ from an infinitely thin disc of diameter
$\sigma$, giving the particle a total diameter $D = \sigma + L$ and
thickness $L$. Note that an OHSC particle, comprising a flat cylindrical
core and a toroidal rim, has a uniaxial symmetry, and its orientation can
be described by a unit vector normal to the central disc,
$\hat{\mathbf{e}}$. The aspect ratio of the OHSC particle is then given
by $L^{*}=L/D$. The pair interaction between two OHSC particles $i$ and
$j$, with respective positions of the center of mass ${\mathbf r}_{i}$ and
${\mathbf r}_{j}$ and orientations $\hat{\mathbf{e}}_{i}$ and
$\hat{\mathbf{e}}_{j}$, is infinite if the shortest
distance between their central discs is less than $L$, and zero
otherwise:
\begin{equation}
    v^{\text{ohsc}}_{ij}(\mathbf{r}_{ij},\hat{\mathbf{e}}_{i},\hat{\mathbf{e}}_{j}) =
    \begin{cases}
        \infty &\text{ if }d_{ij}<L\\
        0 &\text{ otherwise},
    \end{cases}
\end{equation}
where $\mathbf{r}_{ij} = \mathbf{r}_{i} - \mathbf{r}_{j}$ and $d_{ij}$
is the shortest distance between the central discs for particles $i$ and
$j$. We compute this shortest distance using the algorithm outlined in
Ref.~\citenum{cuetos2008columnar}.

We model the interactions between the circular patches by adapting the
Kern-Frenkel potential,~\cite{kern03} where the interactions between a
pair of circular patches are described by a square-well attraction
modulated by an angular factor corresponding to the relative
orientations between the patches. The angular factor is unity only when
the patches are oriented such that the vector connecting the centers of
the two particles passes through both the patches on their surfaces, and
zero otherwise. The width of the square well, $\delta_{\text{cap}}$,
determines the range of the attraction between the patches relative to
the particle diameter. The depth of the square well,
$\varepsilon_{\text{cap}}$, governs the strength of the attractions. The
size of the patches is characterized by a half-angle $\theta$. An
additional parameter $\varphi$ defines the inclination of the plane that
contains the centers of the patches to the plane of the central
cylindrical core.

The total pair potential defining capsomer-capsomer interactions is then
\begin{equation}
  \begin{split}
    v^{\text{cap}}_{ij}(\mathbf{r}_{ij},\hat{\mathbf{e}}_{i},\hat{\mathbf{e}}_{j})
    =&
       v^{\text{ohsc}}_{ij}(\mathbf{r}_{ij},\hat{\mathbf{e}}_{i},\hat{\mathbf{e}}_{j})
       + \\
     &\sum^{5}_{\alpha,\beta} v^{\text{sw,cap}}_{\alpha\beta}(\mathbf{r}_{\alpha\beta})f(\mathbf{r}_{\alpha\beta},\hat{\mathbf{n}}_{i,\alpha},\hat{\mathbf{n}}_{j,\beta}),
  \end{split}
\end{equation}
where $r_{ij}=|\mathbf{r}_{ij}|$ is the center-to-center distance
between particles $i$ and $j$, $\hat{\mathbf{n}}_{i,\alpha}$ is a unit
vector defining the orientation of patch $\alpha$ on particle $i$
(similarly, $\hat{\mathbf{n}}_{j,\beta}$ is a unit vector corresponding to
patch $\beta$ on particle $j$), and $\mathbf{r}_{\alpha\beta}$ is the
separation vector between the centers of patches $\alpha$ and $\beta$.

The term $v^{\text{sw,cap}}_{\alpha\beta}$ is a square-well potential:
\begin{equation}
    v^{\text{sw,cap}}_{\alpha\beta}(r_{\alpha\beta}) =
    \begin{cases}
        -\varepsilon_{\text{cap}} &\text{ if }r_{\alpha\beta}\leq(1+\delta_{\text{cap}})\sigma\\
        0 &\text{ otherwise},
    \end{cases}
\end{equation}
and
$f(\mathbf{r}_{\alpha\beta},\hat{\mathbf{n}}_{i,\alpha},\hat{\mathbf{n}}_{j,\beta})$
is the angular modulation factor,
\begin{equation}
    f(\mathbf{r}_{\alpha\beta},\hat{\mathbf{n}}_{i,\alpha}, \hat{\mathbf{n}}_{j,\beta}) =
    \begin{cases}
      1&\text{if }\hat{\mathbf{n}}_{i,\alpha}\cdot\hat{\mathbf{r}}_{\alpha\beta}>\cos\theta\\
      &\text{ and }\hat{\mathbf{n}}_{j,\beta}\cdot\hat{\mathbf{r}}_{\beta\alpha}>\cos\theta
    \\
    0&\text{otherwise}.
    \end{cases}
\end{equation}

The reference orientation of particle $i$ is such that the normal to the
flat face of the oblate spherocylinder is aligned with the $z$-axis of
the global coordinate frame. We then define the reference position of
the first patch on particle $i$ as $\mathbf{p}_{i,1}=(\sigma/2,0,0)$ and
the position of each other patch as a rotation about the $z$-axis of the
local coordinate frame of the particle such that
$\mathbf{p}_{i,n}=\mathbf{R}_{\psi}\cdot\mathbf{p}_{1}$, where
$\mathbf{R}_{\psi}$ is a rotation matrix defining a clockwise rotation
of angle $\psi=n2\pi/5$ about $\hat{\mathbf{e}}_{i}$ with $n=2,3,4,5$.
The orientation of patch $\alpha$ on particle $i$ is then
$\hat{\mathbf{n}}_{i,\alpha}=\sin(\varphi)\hat{\mathbf{e}}_{i}+(2\cos(\varphi)/\sigma)\mathbf{p}_{\alpha}$,
where $\varphi$ is the angle between $\hat{\mathbf{n}}_{i,\alpha}$ and
the plane containing the flat face of the oblate spherocylinder.

\subsubsection{Polymer-Polymer Interactions.}
Each RNA molecule is modeled as a flexible self-avoiding polymer -- that
is, as a chain of hard-spheres, where neighboring beads in the chain are
connected by a harmonic spring:~\cite{elrad2010encapsulation,
  kampmann2015monte, joseph2021thermodynamics}
\begin{equation}
    v_{\text{poly}}(r_{ij}) = \kappa(r_{ij}-\sigma_{b}l_{b})^{2},
\end{equation}
where $r_{ij}$ is the distance between beads $i$ and $j$ (where
$j=i-1,i+1$), $\kappa$ sets the strength of the harmonic spring,
$\sigma_{b}$ is the hard-sphere diameter of the beads in the polymer
chain, and $l_{b}$ is a dimensionless parameter setting the equilibrium
bond length between neighboring beads.

\subsubsection{Capsomer-Polymer Interactions.}
We allow for interaction between the capsomers and the polymer
\textit{via} an attractive patch on the surface of the capsomer.  The
orientation of the patch is aligned with that of the oblate
spherocylinder. The beads of the polymer and the capsomer then interact
\textit{via} an attractive square-well interaction, plus a hard-core
repulsion between their respective cores. The pair interaction when
particle $i$ is a capsomer and particle $j$ is a bead of a
polymer chain is
\begin{equation}
    v^{\text{cap-pol}}_{ij}(\mathbf{r}_{ij},\hat{\mathbf{e}}_{i}) = v^{\text{hc}}_{ij}(\mathbf{r}_{ij},
    \hat{\mathbf{e}}_{i}) +  v^{\text{sw,cap-pol}}_{ij}(\mathbf{r}_{ij})g(\mathbf{r}_{ij},\hat{\mathbf{e}}_{i}),
\end{equation}
where $v^{\text{hc}}_{ij}$ is the hard-core interaction
\begin{equation}
    v^{\text{hc}}_{ij}(\mathbf{r}_{ij},\hat{\mathbf{e}}_{i}) =
    \begin{cases}
        \infty &\text{ if }d_{ij}<(L+\sigma_{b})/2\\
        0 &\text{ otherwise},
    \end{cases}
\end{equation}
where $d_{ij}$ is the shortest distance between the capsomer and polymer
bead. We compute this distance by first computing the projection of the
polymer bead onto the plane spanned by the cylindrical core of the
capsomer:
$\mathbf{r}^{\text{proj},i}_{j}=\mathbf{r}_{ij}-(\mathbf{r}_{ij}\cdot
\hat{\mathbf{e}}_{i})\hat{\mathbf{e}}_{i}$. Then if
$r^{\text{proj},i}_{j}\leq\sigma/2$, the bead lies over the cylindrical
core of the capsomer, so the shortest distance vector between the two
particles is
$\mathbf{d}_{ij}=\mathbf{r}_{ij}-\mathbf{r}^{\text{proj},i}_{j}$.
Otherwise, the closest point of the capsomer to the bead lies on its
edge. The shortest distance vector between the two particles is then
$\mathbf{d}_{ij}=\mathbf{r}_{ij}-(\sigma/2)\hat{\mathbf{r}}^{\text{proj},i}_{j}$.

The term $v^{\text{sw,cap-pol}}_{ij}$ is the square-well interaction
between the patch on the face of the capsomer and the polymer bead:
\begin{equation}
    v^{\text{sw,cap-pol}}_{ij}(\mathbf{d}_{ij}) =
    \begin{cases}
        -\varepsilon_{\text{cap-pol}} &\text{ if }d_{ij}\leq(1+\delta_{\text{cap-pol}})\sigma \\
        0 &\text{ otherwise}
    \end{cases},
\end{equation}
and $g(\mathbf{r}_{ij},\hat{\mathbf{e}}_{i})$ is the angular modulation
factor for the attractive capsomer-polymer interaction:
\begin{equation}
    g(\mathbf{r}_{ij},\hat{\mathbf{e}}_{i}) =
    \begin{cases}
    1&\text{if } \cos^{-1}(\mathbf{r}_{ij}\cdot\hat{\mathbf{e}}_{i}/r_{ij}) < \pi/2
    \\
    0&\text{otherwise}.
    \end{cases}
\end{equation}

\subsection{Monte Carlo simulations}
We carry out two sets of Monte Carlo simulations in the \textit{NVT}
ensemble using the model outlined above. For the simulations presented
in Fig.~\ref{fig:simulation} we set the volume to be
$V=200000\sigma^{3}$ and the number of polymer chains
$N_{\text{poly}}=10$, with each polymer chain consisting of
$l_{\text{poly}}=150$ beads. For simulations with low capsomer
concentration there are $N_{\text{cap}}=120$ capsomers, for medium
concentration there are $N_{\text{cap}}=300$ capsomers, and for high
concentration there are $N_{\text{cap}}=500$ capsomers. For the larger
simulations containing $N_{\text{poly}}=30$ polymer chains, we set the
volume to be $V=600000\sigma^{3}$, with each polymer chain consisting of
$l_{\text{poly}}=150$ beads. For simulations with low capsomer
concentration simulation there are $N_{\text{cap}}=360$ capsomers, and
in the other simulation there are $N_{\text{cap}}=1500$ capsomers.

We set parameters as follows. We take $\sigma$ to be the unit of length
and $\varepsilon_{\text{cap}}$ to be the unit of energy. We then choose
the parameters defining the system to be $L=0.5\sigma$,
$\delta_{\text{cap}}=0.2$, $\theta=25^{\circ}$, $\varphi=25^{\circ}$,
$\kappa=100\varepsilon_{\text{cap}}$, $\sigma_{b}=0.2\sigma$,
$l_{b}=1.05\sigma_{b}$, and $\delta_{\text{cap-pol}}=0.3\sigma$. For
simulations in which capsids nucleate and grow at low capsomer
concentrations we set
$\varepsilon_{\text{cap-pol}}=0.2\varepsilon_{\text{cap}}$ and the
reduced temperature $k_{\text{B}}T/\varepsilon_{\text{cap}}=0.12$ (where
$k_{\text{B}}$ is the Boltzmann constant, which is taken to be equal to
one), while for simulations in which capsids assemble en masse at low
capsomer concentrations we set
$\varepsilon_{\text{cap-pol}}=0.25\varepsilon_{\text{cap}}$ and
$k_{\text{B}}T/\varepsilon_{\text{cap}}=0.14$. We choose the geometry of the
patches on the capsomers to ensure that the particles can
stabilize a capsid-like structure in which 12 subunits are fully connected
and sit on the vertices of an icosahedron. The choice of the aspect
ratio of the OHSC particles ensures that the cavity of a properly formed
capsid can accommodate cargo of a reasonable size. In turn, the length
of each polymer chain is chosen to be as long as possible with the
constraint that it still fit inside a capsid made of 12 capsomers.

We carry out all Monte Carlo simulations with systems contained in a
cubic box under periodic boundary conditions, using the minimum image
convention. Each capsomer is treated as a rigid body for which the
orientational degrees of freedom are represented by quaternions. The
potential energy is calculated using a spherical cutoff of $1.7\sigma$,
and a cell list is used for efficiency. Each Monte Carlo cycle consists
of $N$ translational or rotational single-particle or cluster moves,
chosen at random with equal probabilities.

\section*{Author contributions}
All authors conceived the research goals and aims. LAW and AN curated
data and code. LAW and AN analyzed the results, with assistance from VNM
and DC. RFG, DC, and VNM obtained funding. LAW performed the experiments
and AN performed the simulations. All authors developed methods. RFG,
DC, and VNM administered the project. AN developed the computer software
for the simulations. RFG, DC, and VNM supervised the project. LAW and AN
performed validation studies. LAW and AN prepared visualizations, edited
by VNM. LAW, AN, and VNM prepared the original draft. All authors
reviewed and edited the submitted work.

\section*{Conflicts of interest}
There are no conflicts to declare.

\section*{Data availability statement}
Experimental data are freely available at the Harvard
Dataverse~\cite{DVN/8A2HWD_2023}. Simulation code and results are
available at the UBIRA eData repository.

\section*{Acknowledgements}

We thank Amy Barker and Peter Stockley at the University of Leeds for
initial stocks of MS2 and \textit{E.\@ coli} cells. We thank Tim Chiang,
Amelia Paine, Aaron Goldfain, and Danai Montalvan for helpful scientific
discussions. This research was partially supported by a National Science
Foundation (NSF) Graduate Research Fellowship under grant number
DGE-1745303, by NSF through the Harvard University Materials Research
Science and Engineering Center under NSF grant number DMR-2011754, by
the National Institute of General Medical Sciences of the National
Institutes of Health under grant numbers K99GM127751 and R00GM127751, by
the NSF-Simons Center for Mathematical and Statistical Analysis of
Biology at Harvard University under NSF grant number 1764269, and by the
Harvard Quantitative Biology Initiative. AN, VNM, and DC gratefully
acknowledge support from the Institute of Advanced Studies of the
University of Birmingham and the Turing Scheme. This work was performed
in part at the Harvard University Center for Nanoscale Systems (CNS), a
member of the National Nanotechnology Coordinated Infrastructure Network
(NNCI), which is supported by the National Science Foundation under NSF
grant number ECCS-2025158. The work was also performed in part at the
Harvard University Bauer Core Facility. Any opinion, findings, and
conclusions or recommendations expressed in this material are those of
the authors and do not necessarily reflect the views of the National
Science Foundation.




\renewcommand\refname{References}
\bibliographystyle{rsc} 
\bibliography{bibliography}

\clearpage

\pagebreak
\renewcommand\thefigure{S\arabic{figure}}
\renewcommand\thepage{S\arabic{page}}
\setcounter{figure}{0}
\setcounter{page}{1}

\fancyfoot{}
\fancyfoot[C]{\footnotesize{\sffamily{\thepage}}}
\fancyhead{}

\twocolumn[
\begin{@twocolumnfalse}
  \sffamily

  \vspace{3in}
  \begin{center}
    \noindent\LARGE{\textbf{Supplementary Information for}} \\
    \noindent\LARGE{\textbf{Effect of coat-protein concentration on the self-assembly of bacteriophage MS2 capsids around RNA}} \\
 
 \noindent\large{LaNell A. Williams,\textit{$^{a}$}
 Andreas Neophytou,\textit{$^{b}$}
 Rees F. Garmann,\textit{$^{c,d,e}$}
 Dwaipayan Chakrabarti,\textit{$^{b}$}
   Vinothan N. Manoharan,$^{\ast}$\textit{$^{c,a}$}}\\

\end{center}

\end{@twocolumnfalse} \vspace{0.6cm}

]
\pagebreak

\begin{figure*}[htp] 
  \centering
  \includegraphics[width=\textwidth]{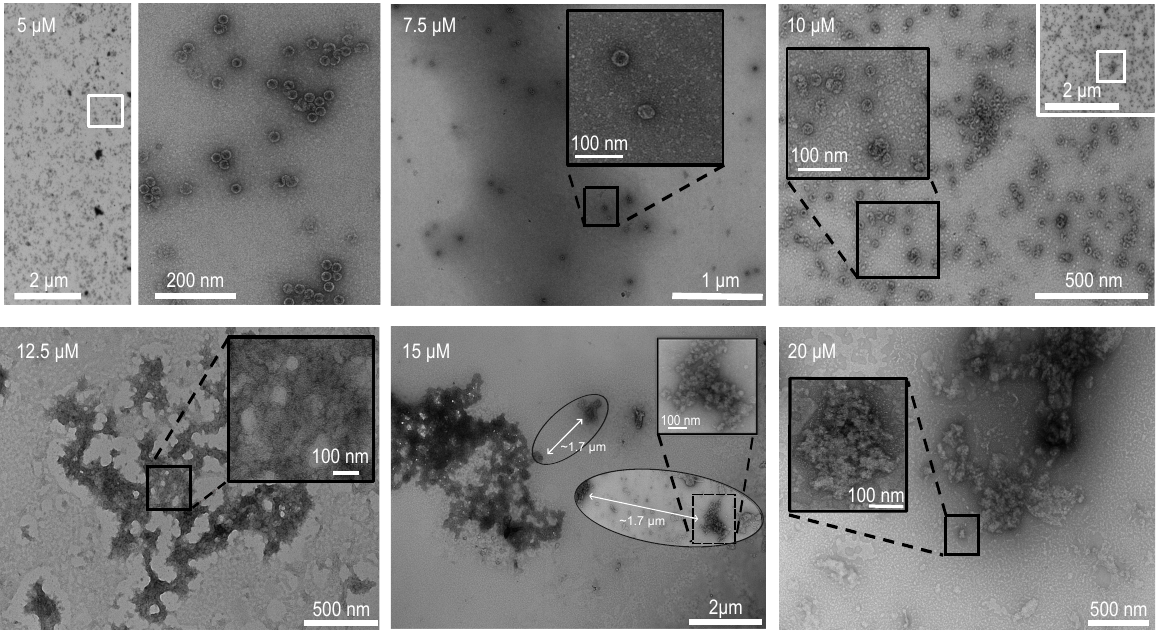}
  \caption{\label{fig:farviewcondensates_revised_300dpi} TEM images of
    the same samples as shown in
    Fig.~\ref{fig:temfig_editable_revised_300dpi}, but at different
    magnifications for each sample (some rotation and translation may be
    present between the low-magnification images and the insets due to drift in
    the microscope). The low magnification images of the samples at 12.5
    $\mu$M coat-protein dimer concentration and higher show
    micrometer-sized structures.}
\end{figure*}

\begin{figure*}[hbp]
  \centering
  \includegraphics[width=12cm]{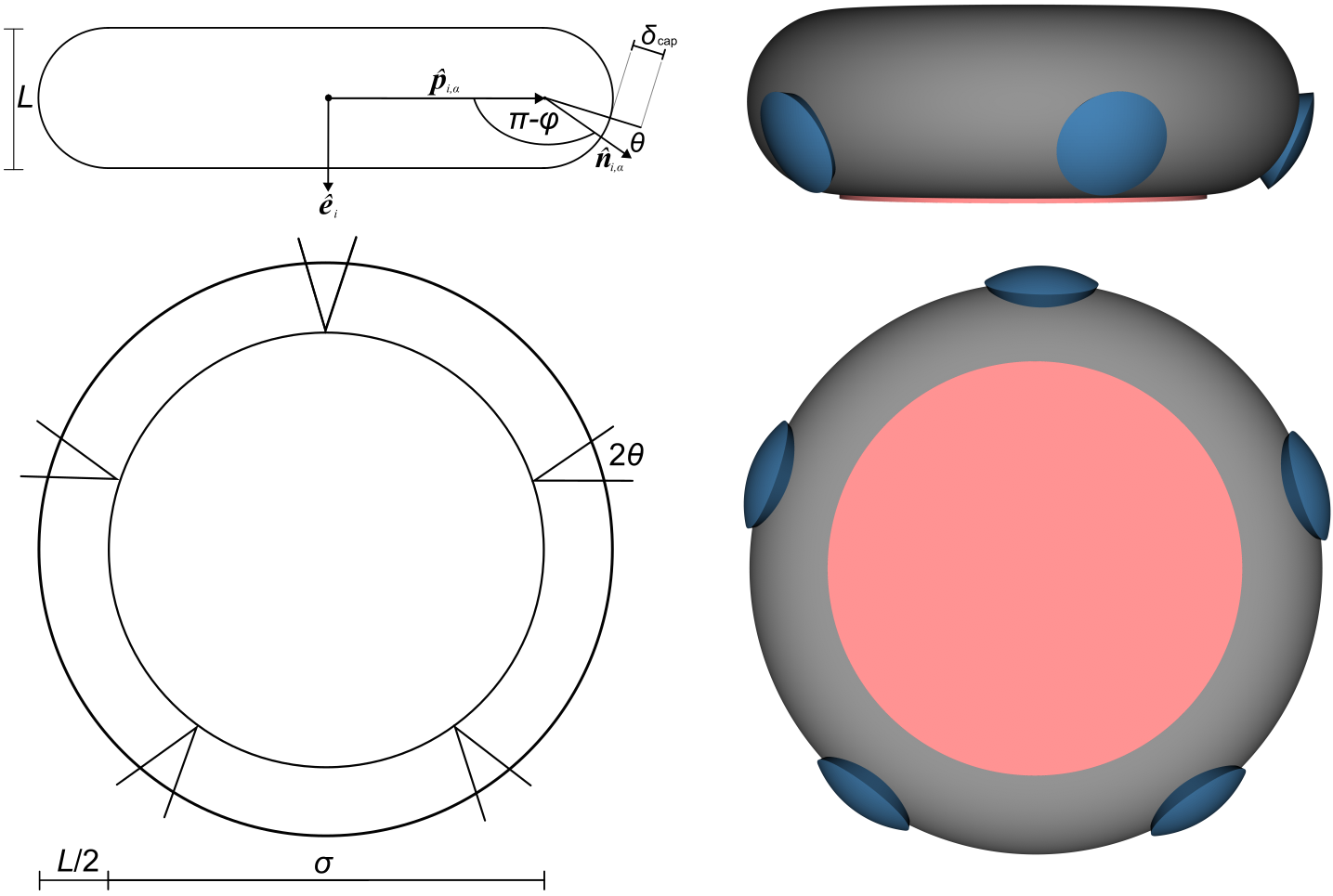}
  \caption{\label{fig:patchymodel} Schematic (left) and
    three-dimensional (right) representation of the patchy particle
    model for capsomers. The schematic view of the model shows key model
    parameters that define the geometry of the hard core of each
    capsomer particles, namely the diameter $\sigma$ and thickness
    $L$.  Additionally, parameters that define the geometry of
    the attractive patches are shown for one of the five patchy sites,
    where $\hat{\mathbf{p}}_{i,\alpha}$ defines the position of the
    patch in the local coordinate frame of the capsomer,
    $\hat{\mathbf{n}}_{i,\alpha}$ is the vector defining the orientation
    of the patch, $\theta$ is the half-angle of the patch, $\varphi$ is
    the angle between $\hat{\mathbf{n}}_{i,\alpha}$ and the plane
    containing the circular core of the capsomer and
    $\delta_{\text{cap}}$ is the range of the attractive patch-patch
    interactions.  We show a three-dimensional representation of
    the particles used in this work on the right.
  }
\end{figure*}

\begin{figure*}[hbp]
  \centering
  \includegraphics[width=16cm]{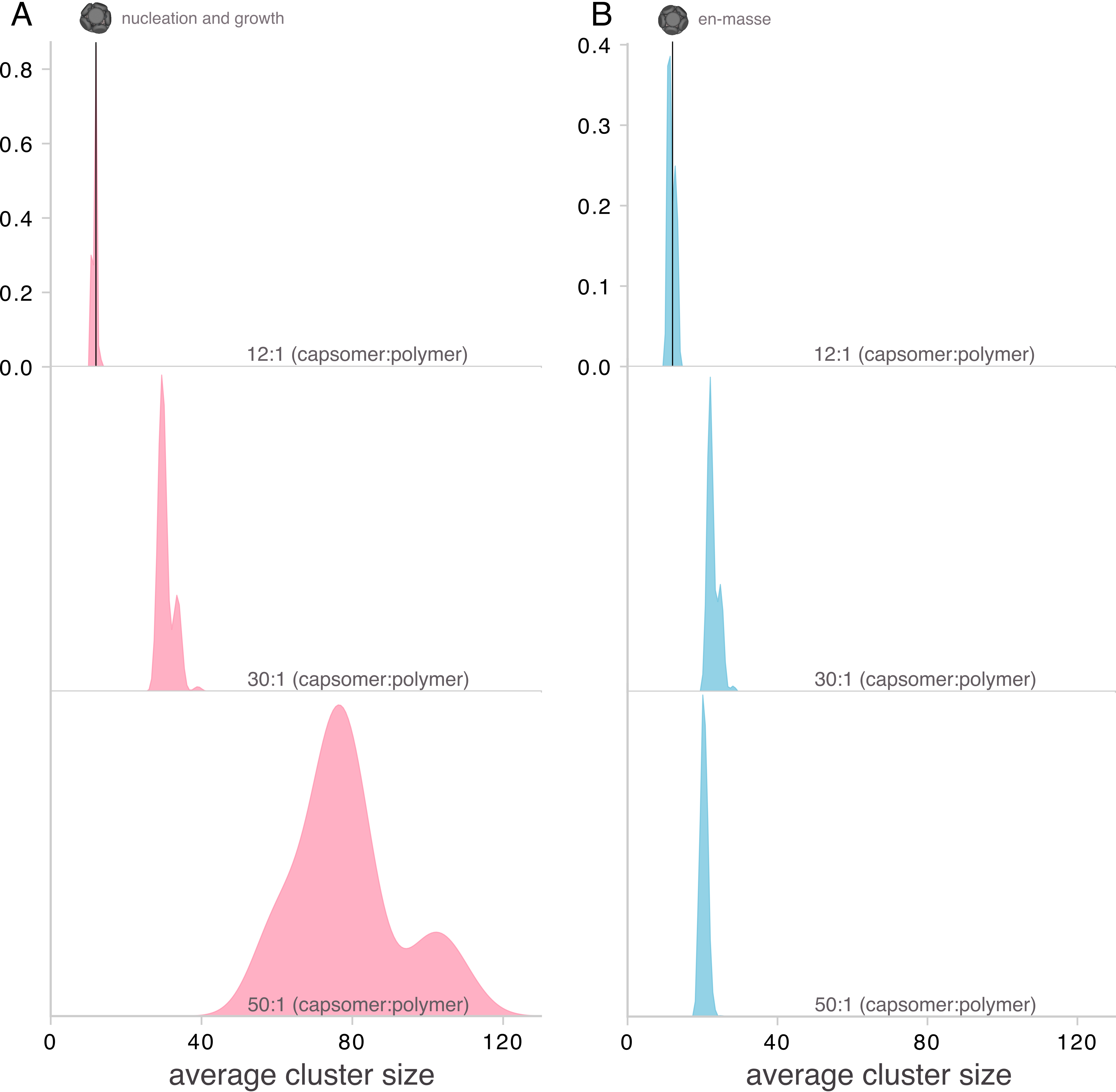}
  \caption{\label{fig:Simulation_DLS} Average cluster size distributions
    observed in Monte Carlo simulations of capsomer-and-polymer systems
    where the relative energies of capsomer-capsomer and
    capsomer-polymer interactions are chosen such that capsid formation
    in the low capsomer:polymer ratio systems proceeds either by (A) a
    nucleation and growth pathway or (B) an en-masse pathway. For both
    sets of simulations, the average cluster sizes are computed for
    systems containing 12:1, 30:1 and 50:1 ratio of capsomer to polymer.
  }
\end{figure*}

\begin{figure*}[hbp]
  \centering
  \includegraphics[width=16cm]{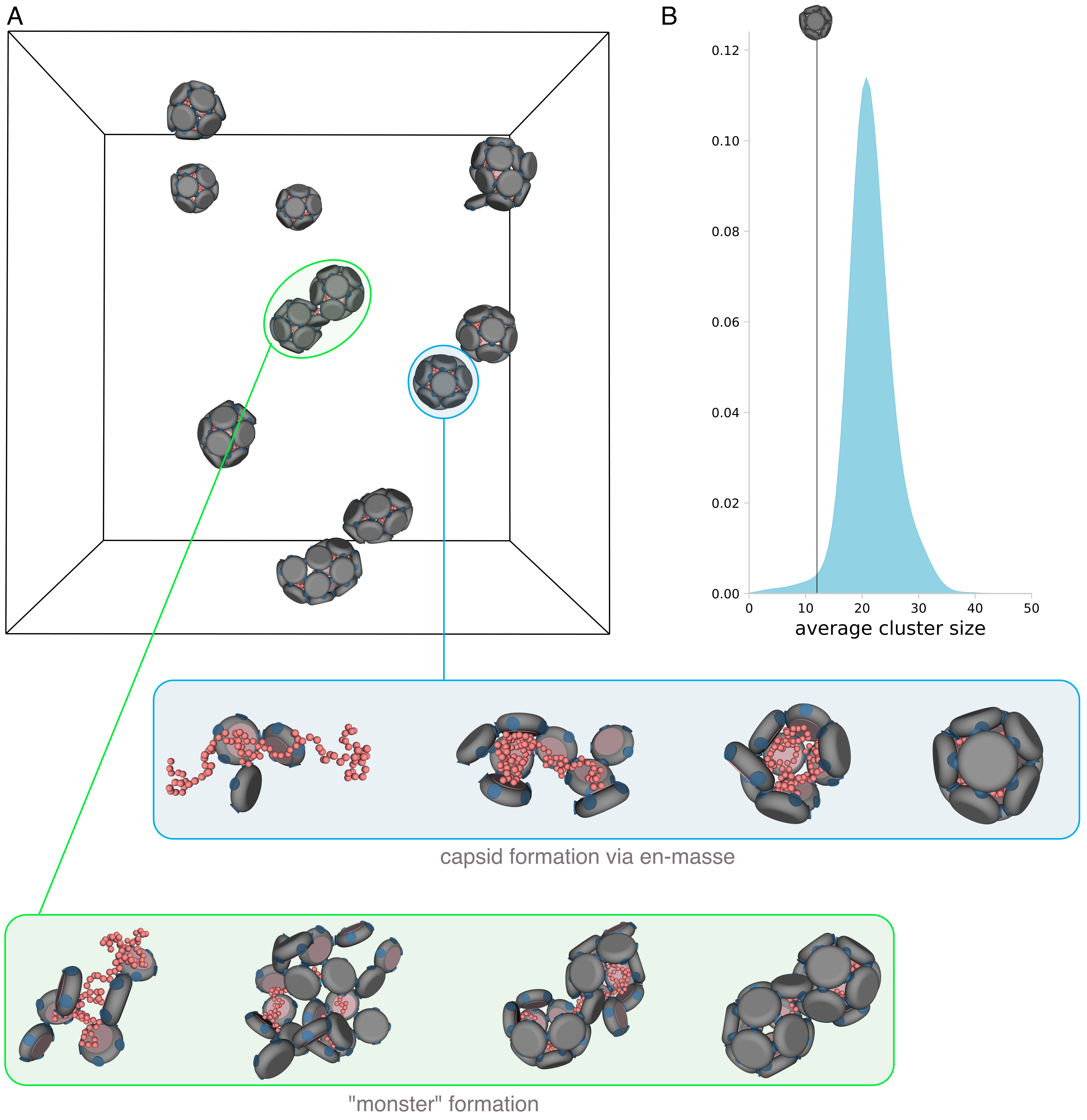}
  \caption{\label{fig:Simulation_short_poly}
    Results of Monte Carlo simulations at a high capsomer:polymer ratio
(50:1) with short polymer chains and capsomer-capsomer and
capsomer-polymer interactions chosen such that capsids nucleate and grow
in the low capsomer:polymer ratio systems.
    (A) Representative snapshot of clusters that
contain polymer chains. Two clusters have been highlighted. One
is a monster particle that forms when two capsid shells
assemble around a single polymer chain, and the other is a capsid that
assembles \textit{via} an en-masse pathway, as shown in the shaded boxes.
    (B) Average cluster size distribution for the system, showing that
when the polymer chains are short, the system contains no condensates,
and only capsids and monsters form.
  }
\end{figure*}

\clearpage

\end{document}